\documentclass [twocolumn,showpacs,prl]{revtex4}
\usepackage{graphicx}
\usepackage{bm}
\begin{document}
\narrowtext
\title{Laser-Generated Ultrashort Multi-Megagauss Magnetic Pulses in Plasmas}   
\author{A. S. Sandhu$^1$, A. K. Dharmadhikari$^1$, P. P. Rajeev$^1$, G. R. Kumar$^1$, S. Sengupta$^2$, A. Das$^2$, and P. K. Kaw$^2$}
\address{$^1$Tata Institute of Fundamental Research, 1 Homi Bhabha Road, Mumbai 400 005, India}
\address{$^2$Institute for Plasma Research, Bhat, Gandhinagar, Ahmedabad 382428, India}
\date{\today}

\begin{abstract}
	We demonstrate ultrashort (6 ps), multi-Megagauss (27 MG) magnetic pulses generated upon interaction of an intense laser pulse $(10^{16}$ Wcm$^{-2}$, 100 fs) with a solid target. The temporal evolution of these giant fields generated near the {\it high density} critical layer is obtained with the highest resolution reported so far. Particle-in-cell simulations and phenomenological modeling is used to explain the results. The first direct observations of anomalously rapid damping of plasma shielding currents produced in response to the hot electron currents penetrating the bulk plasma are presented. 

\end{abstract}
\maketitle

	The largest magnetic fields available terrestrially ($\sim 10^{8}$ Gauss) are generated by explosive ionization of a solid target with an intense ultrashort laser pulse \cite{Tatarakis}. Since the first observation of such magnetic fields, their origin, magnitudes and other qualitative features have attracted considerable attention \cite{Stamper}. Recently, sub-picosecond laser produced solid plasmas have provided a new experimental facet to these studies. Magnetic fields up to gigagauss magnitudes have been predicted in overdense region of solid target \cite{Sudan}.  Little, however, is known about the temporal evolution of these huge magnetic fields generated around critical layer. These fields play a crucial role in electron transport \cite{Mason} and are, therefore, important for potential applications in hybrid confinement \cite{Kolka} and fast ignition \cite{Tabak} schemes of laser fusion.  

	In this Letter, we present first experimental measurements of temporal evolution of Megagauss magnetic fields generated {\it at the critical layer}, on femtosecond timescales. The field generation and decay mechanisms are identified and role of resonance absorption is examined. The initial buildup of magnetic field due to direct laser radiation effects is calculated using LPIC++ code \cite{LPIC} and the results are found to be in good agreement with the experiment. The field evolution is explained to be due to currents generated by fast electrons \cite{Sentoku} and plasma return currents damped by turbulence induced resistivity \cite{Drake}. The first direct observation of anomalously rapid damping of return currents, which may have important consequences for laser fusion, is reported. We demonstrate ultrashort, Megagauss magnetic pulses, with 6 picosecond (FWHM) duration and a peak magnitude of 27 Megagauss generated by a p-polarized laser pulse.  Possible applications of such high-field ultrashort magnetic pulses in various areas are discussed.

	Laser pulses (806 nm, 100 fs) are derived from a custom built chirped pulse amplification Titanium:Sapphire laser, described in detail elsewhere \cite{Banerjee}. The linearly polarized pump pulse incident at $55^{\circ}$ is focused on a solid target with typical intensities of $10^{16}$ W cm$^{-2}$ (Fig. 1). A small part of the laser is used to generate a linearly polarized, second harmonic (403nm) probe pulse using a thin (0.5mm) BBO crystal. The probe pulse incident at $50^{\circ}$, is focused to an intensity of $ 5 \times 10^{12}$ W cm$^{-2}$ and is spatially overlapped with the pump spot. The probe penetrates beyond the critical density($n_{c}$) for pump (i.e. upto 1.65$n^{806}_{c}$ or 0.4$n^{403}_{c}$ as shown in inset of Fig.1). The probe pulse delay with respect to pump is varied using a motorized translation stage. 

	The novel features of this experiment are:  (a) we are able to examine the generated magnetic field on either side of the critical layer by using the second harmonic of the laser as a non-tangential probe, (b) we identify zero-delay time precisely and obtain high-resolution data for temporal evolution, and (c) we examine, using oblique pump incidence, the role of Resonance Absorption (RA). To our knowledge, only one laboratory has hitherto reported \cite{Borghesi} pump-probe magnetic fields measurements on picosecond time scales (using ~1 ps duration laser pulse at normal incidence). However, these measurements were confined to an under-dense plasma region $(n_e = 4\times 10^{19}$ cm$^{-3})$ and the diagnostics employed limited the temporal resolution to $ \sim 3$ ps. We emphasize our probing technique, because, magnetic field generation primarily occurs near the critical surface (electron density $ \sim 10^{21} cm^{-3}$), the region of maximum laser absorption \cite{Sentoku}.  Moreover, it is the magnetic field in the over-dense region that determines hot electron transport into the bulk, which is crucial for fusion related issues. 

\begin{figure}
\centering
\includegraphics [width=2.8in,height=3in]{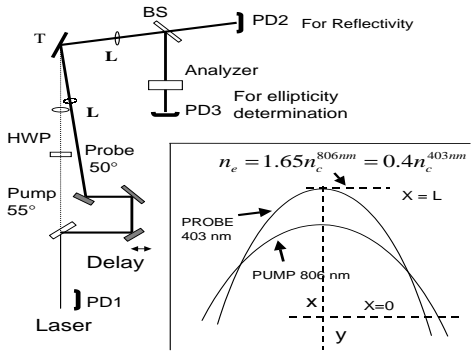}
\caption{The experimental setup, Target (T), Lens (L), Beam Splitter (BS), Photo Diode (PD), Half Wave Plate (HWP). Inset shows diagrammatic representation of pump and probe laser paths in the plasma. The critical densities for pump and probe are given as $n^{806nm}_c$ and $n^{403nm}_c$ respectively}
\end{figure}

	We use standard Cotton-Mouton polarimetry \cite{Hutchinson}, which involves measurement of magnetically induced ellipticity in the polarization state of the probe pulse. The contributions to induced ellipticity from plasma propagation and refraction effects \cite{Ginzburg} are calculated and are found to be insignificant \cite{Calculation}. The targets used in these experiments are optically polished aluminum discs. The target is housed inside a vacuum chamber at $10^{-3}$ Torr and is translated after each laser shot, exposing a fresh surface for interaction. The pump and probe focussing is monitored using hard x-ray emission as diagnostic. The spot radii for pump and probe are 15$\mu$m and 10$\mu$m respectively. The reflected probe beam is collected using f/20 lens. The amplitude and polarization of the specularly reflected probe is simultaneously monitored by splitting it into two arms, one measuring the reflectivity with a photodiode (PD2) and the other measuring the polarization state using an analyzer (extinction ratio $10^{-5}$) in front of another identical photodiode (PD3), as shown in Fig. 1. The photodiode PD1 measures the shot to shot laser fluctuations. All three PD signals are simultaneously recorded for each laser shot along with the delay stage position. The ellipticity is determined at various fixed delay positions by plotting PD3 signal with respect to analyzer angle. Higher temporal resolution ellipticity data is obtained by continuously varying the delay in steps of 1$\mu m$ at different fixed analyzer positions and computing the ratio PD3/PD2 for each delay. The ratio (PD3/PD2) is used above so as to account for plasma reflectivity variation as function of time delay. The temporal behavior of the magnetic field from -2 to +10 ps time delay is monitored. The constant background noise due to the second harmonic radiation from the plasma generated by the pump pulse is negligible ($<0.1 \%$) as compared to the reflected probe.

	Figure 2 presents the temporal evolution of the magnetic field. The inset shows reflectivity and induced ellipticity of the probe for a p-polarized pump. The sharp reflectivity dip is used to independently establish the start of the magnetic pulse.  The magnetic field is derived from induced ellipticity ($\beta$) for our experimental conditions using \cite{Hutchinson} $\beta (t) = 3.32 \times 10^{-26} \int n_{e}(l,t) B^{2}(l,t) dl$, where `$n_e$' is the electron density in $cm^{-3}$, `$B$' is magnetic field in MG, and `$l$' is the path length in $\mu$m. The magnetic field is deduced assuming a spatially uniform $B$ over a linear density gradient $n_{e}(x) = 0.4n^{403}_{c}(x/L)$, where $L$ is the plasma slab length, as shown in the inset of fig1. The factor $0.4n^{403}_c$ corresponds to the turning point density for the 403nm probe. We integrate over the trajectory in the plasma, $dl = \sqrt{1+y'^{2}}dx$, where $y'= sin{\theta}_{0}/\sqrt{\epsilon (x)-sin^{2}\theta_0}$ , $\epsilon(x)$ being the dielectric function. The plasma expansion velocity is estimated to be $5 \times 10^{7}$ cm/s from Doppler shift measurements of the reflected probe.  With $L(t)= 1+ v_{exp} t$ ,  the magnetic field as a function of time delay is obtained as $B(t)= 80 \sqrt{\beta (t)/(1+0.5t)}$ MG  . From the results shown in Fig. 2, the magnetic field pulse generated p-polarized pump has a peak value of 27 MG and duration (FWHM) of 6ps.  In comparison s-polarized pump results in a peak value of 14 MG. This confirms the importance of RA, induced by p-polarization, in magnetic field generation. The magnetic field in case of s-polarized pump can be explained to be arising from parametric instabilities near critical density, which develop at a slower rate as seen in figure 2. Further, any realistic laser focusing always involves some RA contribution even for s-polarization, due to critical surface rippling and geometrical effects \cite{Sentoku,Rajeev}.

\begin{figure}
\includegraphics [width=2.8in,height=3in]{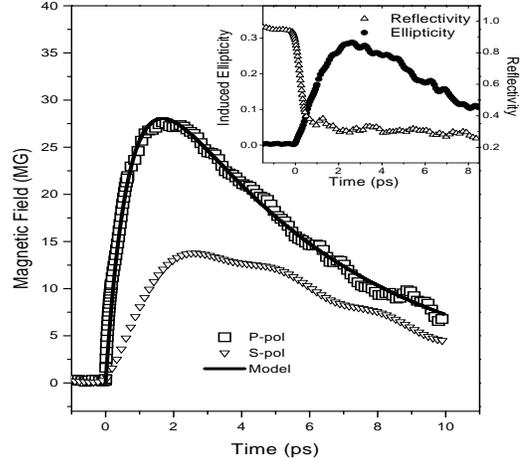}
\caption{Magnetic field pulse profile for p- and s-polarized pump (open squares and triangles respectively) laser with intensity of $1.1 \times 10^{16}$ W cm$^{-2}$. Solid line shows the fit obtained in case of p-polarized pump using our phenomenological model. The inset shows the reflectivity and induced ellipticity of probe as function of delay time.}
\end{figure}

	The mechanism of quasi-static magnetic field generation in ultrashort laser -plasma experiments can be understood by invoking the following magnetic field evolution equation

\begin{equation}
\frac{\partial \vec{B}}{\partial t} = - \vec{\nabla} \times (\frac{\vec{j}
\times \vec{B}}{n_{e}e}) + \frac{c}{n_{e}e} ( \vec{\nabla}T_{e} \times
\vec{\nabla}n_{e} ) + \frac{c}{\sigma} ( \vec{\nabla} \times \vec{j}_{hot}
) + \frac{c^{2}}{4 \pi \sigma} \nabla^{2} \vec{B},
\end{equation}

Equation (1) is derived by taking the curl of the equation of motion of background plasma electrons that carry the plasma shielding currents, $\vec{j}_p = \vec{\nabla} \times \vec{B} - \vec{j}_{hot}$ . The first term in equation (1) is the standard electron magnetohydrodynamic (EMHD)\cite{Kingsep} source due to the Hall effect, the second term is the thermoelectric source and the third term is the source due to hot electrons (generated near the pump critical surface by the RA of the laser pulse); the last term gives the magnetic field decay due to resistive damping of the plasma shielding currents ($\sigma^{-1}$ being the background plasma resistivity). Electron inertia and magnetic field convection effects are assumed to be negligible in equation (1).

We estimate the first term (important during the first 200 fs when the pump is on) by taking a cycle-averaged product of high frequency current and magnetic fields \cite{DeGroot},$ \vec{\nabla}\times \left<(\vec{j_{h}} \times \vec{B_{h}})/n_{e}e\right>$, inside the plasma. We obtain the magnitude of this term by carrying out a particle-in-cell simulation using the laser-plasma interaction code LPIC++. In the simulation, a p-polarized light pulse with $sin^{2}$ envelope and 100 fs FWHM laser pulse is incident at an angle of $55^{\circ}$ on a linear density ramp. Fig. 3(a) shows numerically obtained spatial and temporal profiles of the quasi-steady state magnetic field. This confirms the earlier analytical results of spatial profiles of laser generated magnetic fields \cite{Thompson}. At the same time, the values of $B$ and $\partial B / \partial t$ obtained from simulation compare well with our experimental results (Fig. 3b).  The numerically obtained maximum value of $\partial B / \partial t$ is 22 MG/ps at 120 fs, which is close to the experimentally deduced $ \partial B / \partial t$.  The $B$ value reached in 150 fs is 1.2 MG, beyond which the simulation results saturate. This is expected, because to model this regime we have considered only one term, which is valid strictly for the duration of pulse. 

The second term in equation (1) is estimated assuming a temperature gradient of 100 eV over a transverse scale of $15 \mu$m (pump spot radius) and density gradient of $10^{21}$ cm$^{-3}$ over $1 \mu$m in the normal direction. This gives $\partial B/ \partial t \sim 0.05$ MG/ps, which is much smaller than the experimental results. Hence, the thermoelectric source term is neglected in subsequent analysis.

\begin{figure}
\includegraphics [width=2.8in,height=3in]{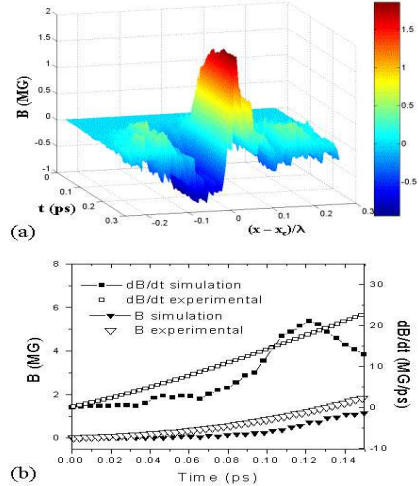}
\caption{(a) The temporal and spatial profile of magnetic field obtained using the LPIC++ code ($x=x_c$ is the critical layer). (b) Comparison of B and dB/dt obtained using LPIC++ simulation with experimental results.}
\end{figure}

The remaining terms (magnetic diffusion and hot electron source \cite{Yablonovitch,Sentoku} ) in the magnetic field evolution equation govern the evolution of $B$ after the pump pulse is removed. We model this regime by a phenomenological 0-dimensional evolution equation 
\begin{equation}
\frac{\partial B}{\partial t} = S(t) - \frac{B}{\tau},
\end{equation} 
where $S(t)$  is the source term. $S(t)$  is mainly due to the fast electron currents and $B/\tau$ is a 0-d representation of the magnetic diffusion term. Taking $S(t) = S_{0} exp( - t / t_{0} )$, we get
\begin{equation}
B(t) = \frac{S_{0}}{1/\tau - 1/t_{0}}[exp(-t/t_{0}) - exp(-t/\tau)],
\end{equation}					
The first term in equation (3) denotes the natural decay of the hot electron source produced by RA mechanism. The second term describes the resistive decay of the fields generated by the plasma return currents. As shown in Fig. 2, this expression gives an excellent fit to our experimental data for p -polarized pump with $S_{0} = 53.7$ MG/ps, $t_{0} = 0.7$ ps and $\tau = 5.6$ ps. 

To get an insight into these numerical values, we estimate conductivity $\sigma$ from the magnetic diffusion term. Using $\tau \sim ( 4 \pi \sigma / c^{2} ) (\Delta x)^{2}$, for the best fit value $\tau$ = 5.6 ps and $\Delta x = 15 \mu$m , we get $\sigma = 1.8 \times 10^{14}$ sec$^{-1} $. However the $\sigma_{solid}$  or $\sigma_{classical}$  observed \cite{Milchberg} at $T_e =100$ eV has value of $\approx 4.5 \times 10^{15}$ sec$^{-1}$, which is an order of magnitude greater than that obtained from magnetic field decay! This clearly brings out the importance of turbulence induced anomalous resistivity effects in the damping of shielding plasma currents. An upper estimate of anomalous resistivity due to turbulence of electrostatic waves, obtained by taking the effective collision frequency $\nu_{eff} \sim f \omega_{p}$ ( where $f$ is a fraction of order unity ), is in reasonable agreement with our estimate of $\sigma$. We note that the hot electron source term is effective for about 0.7 ps, which is longer than the laser pulse (FWHM=100fs). Thus we argue that the electrostatic plasma waves generated by the RA mechanism (typically with $E^{2}/ 4 \pi n_{e} T_{e} \gg 1$ ) during the laser pulse will continue to slowly damp (or convect away) and accelerate electrons for a few hundred femtoseconds after the laser pulse.  
To get an estimate of $S_{0}$, which is the magnitude of the source term $\mid (c/\sigma) (\vec{\nabla} \times \vec{j}_{hot} ) \mid$ , we evaluate $j_{hot} \approx f_{a} e I / T_{hot}$ by taking a conversion fraction ($f_a$) of incident energy into hot electrons as 0.3 and $T_{hot} \sim 20$ keV (estimated by using the well known scaling laws for resonance absorption \cite{Estabrook}). This yields hot electron current density $j_{hot} \sim 4.5 \times 10^{20}$ statampere/cm$^{2}$. Using anomalous conductivity obtained above we get $S_{0} \sim  61$ MG/ps, which is again in close agreement with our phenomenological fit.

In conclusion, we have measured and characterized picosecond megagauss magnetic pulses generated by the interaction of ultrashort laser pulse with a solid. Our measurements extend to overdense region of the target and hence are of relevance to electron transport and fusion related issues. The experimentally observed rise times and magnitude of magnetic fields closely follow theoretical estimates and simulations. We also observe for the first time, anomalously rapid damping of return plasma shielding currents produced in response to the hot electron currents penetrating the bulk plasma; this is a topic of great significance to the fast ignition scheme. Such ultrashort, localized magnetic fields are useful for investigating magnetic precession and reversal dynamics, which is vital for developing next generation ultrafast switching and storage devices \cite{Siegmann}. Further, the generation and characterization of these giant magnetic fields offers a unique opportunity for accessing extreme stellar conditions and testing astrophysical theories in the laboratory \cite{Lai}. The diagnostics in chemical and biological sciences like Magnetic Circular Dichroism and Magnetic Resonance may also benefit in situations where high fields are essential \cite{Que}. The intensities used in our experiments are easily realizable with modern kilohertz repetition rate femtosecond lasers, and we foresee exciting applications for these magnetic pulses.

	We would like to thank B. Bhattacharya and V.K. Tripathi for fruitful discussions, J. Meyer-ter-Vehn and T. Schlegel for the LPIC++ code and  D. Mathur and S. Maiti for useful suggestions. The TIFR high intensity laser system used was partially funded by the Department of Science and Technology.


\begin{references}
\bibitem{Calculation} The effective phase difference is given as, 
$
\delta \phi_{sp}=2(\frac{\omega}{c})\int[\epsilon(x)- sin^{2}\theta]^{1/2} dx - 2(\frac{\omega}{c})\int[\epsilon(x)- sin^{2}\theta +(\frac{c}{\omega})^{2}(\frac{1}{2\epsilon}\frac{d^{2}\epsilon}{dx^{2}}- \frac{3}{4\epsilon^{2}}(\frac{d\epsilon}{dx})^{2})]^{1/2}dx $.
For our parameters, we get maximum phase difference of 0.12 radian. In the worst-case scenario, where the plane of polarization is at $ 45^{\circ}$ to the plasma gradient, we get ellipticity that is below the baseline (i.e. observed when probe alone is present and, hence, is subtracted from the data) value of 0.07.
\end{references}
\end{document}